\documentclass[prd,superscriptaddress,nofootinbib,secnumarabic,preprintnumbers]{revtex4-2}[10pt]
\pdfoutput=1
\usepackage{latexsym}
\usepackage{mathrsfs, amsmath, amssymb}
\usepackage{diagbox, slashed, makecell}
\usepackage{hyperref, url}
\usepackage{color}
\usepackage{graphicx}
\usepackage{appendix}
\usepackage{verbatim}
\usepackage{multirow}
\usepackage{ucs}
\usepackage{indentfirst, layout, geometry}

\newcommand{\mr}{\mathrm}
\newcommand{\mc}{\mathcal}

\graphicspath{{figures/}}

\begin{document}

\begin{sloppypar}

\title{Scalar leptoquark and vector-like quark extended models as the explanation of the muon $g-2$ anomaly: bottom partner chiral enhancement case}

\author{Shi-Ping He}
\email{shiping.he@apctp.org}
\affiliation{Asia Pacific Center for Theoretical Physics, Pohang 37673, Korea}

\date{\today}

\begin{abstract}
Leptoquark (LQ) models are well motivated solutions to the $(g-2)_{\mu}$ anomaly. In the minimal LQ models, only specific representations can lead to the chiral enhancements. For the scalar LQs, the $R_2$ and $S_1$ can lead to the top quark chiral enhancement. For the vector LQs, the $V_2$ and $U_1$ can lead to the bottom quark chiral enhancement. When we consider the LQ and vector-like quark (VLQ) simultaneously, there can be more scenarios. In our previous work, we considered the scalar LQ and VLQ extended models with up-type quark chiral enhancement. Here, we study the scalar LQ and VLQ extended models with down-type quark chiral enhancement. We find two new models with $B$ quark chiral enhancements, which originate from the bottom and bottom partner mixing. Then, we propose new LQ and VLQ search channels under the constraints of $(g-2)_{\mu}$.
\end{abstract}

\maketitle

\clearpage

\section{Introduction}

The $(g-2)_{\mu}$ anomaly is a longstanding puzzle in the standard model (SM) of elementary particle physics. It is first announced by the BNL E821 experiment \cite{Muong-2:2006rrc}. Last year, the FNAL muon $g-2$ experiment reports the increased deviation from the SM prediction \cite{Muong-2:2021ojo}. When combining the BNL and FNAL data, the averaged result is $a_{\mu}^{\mr{Exp}}=116592061(41)\times10^{-11}$. Compared to the SM prediction $a_{\mu}^{\mr{SM}}=116591810(43)\times10^{-11}$ \cite{Aoyama:2012wk, *Aoyama:2019ryr, *Czarnecki:2002nt, *Gnendiger:2013pva, *Davier:2017zfy, *Keshavarzi:2018mgv, *Colangelo:2018mtw, *Hoferichter:2019mqg, *Davier:2019can, *Keshavarzi:2019abf, *Kurz:2014wya, *Melnikov:2003xd, *Masjuan:2017tvw, *Colangelo:2017fiz, *Hoferichter:2018kwz, *Gerardin:2019vio, *Bijnens:2019ghy, *Colangelo:2019uex, *Blum:2019ugy, *Colangelo:2014qya, *Aoyama:2020ynm}, the deviation is $\Delta a_{\mu}\equiv a_{\mu}^{\mr{Exp}}-a_{\mu}^{\mr{SM}}=(251\pm59)\times10^{-11}$, which shows $4.2\sigma$ discrepancy. Then, many new physics models are proposed to explain the anomaly \cite{Czarnecki:2001pv, Jegerlehner:2009ry, Freitas:2014pua, Queiroz:2014zfa, Lindner:2016bgg, Athron:2021iuf}. 

For the mediators with mass above TeV, the chiral enhancements are required, which can show up when left-handed and right-handed muon couples to a heavy fermion simultaneously. In the new lepton extended models \cite{Kannike:2011ng, Frank:2020smf, Dermisek:2021ajd, Crivellin:2021rbq}, the chiral enhancements originate from the heavy lepton mass. Besides, the LQ models can be the alternative choice \cite{Cheung:2001ip, Dorsner:2016wpm, ColuccioLeskow:2016dox, Dorsner:2019itg, Crivellin:2020tsz, Dorsner:2020aaz, Zhang:2021dgl}, in which the chiral enhancements originate from the large quark mass. For the minimal LQ models, there are scalar LQs $R_2/S_1$ with top quark chiral enhancement and vector LQs $V_2/U_1$ with bottom quark chiral enhancement. The LQ can connect the lepton sector and quark sector. On the other hand, the VLQ naturally occurs in many new physics models and is free of quantum anomaly. It can mix with SM quarks and provide new source of CP violation. Hence, the LQ and VLQ extended models can lead to interesting flavour physics in both lepton sector and quark sector. In our previous paper \cite{He:2021yck}, we investigated the scalar LQ and VLQ \footnote{The terminology VLQ should not be confused with the vector leptoquark in some bibliographies.} extended models with top and top partner chiral enhancements. In this work, we will study the scalar LQ and VLQ extended models, which can produce the bottom partner chiral enhancements. This paper is complementary to our paper \cite{He:2021yck}. Moreover, the top partner and bottom partner lead to different collider signatures.

In Sec. \ref{sec:model}, we introduce the models and show the related interactions. Then, we derive the new physics contributions to $(g-2)_{\mu}$ and perform the numerical analysis in Sec. \ref{sec:g-2}. In Sec. \ref{sec:phenomenology}, we discuss the possible collider phenomenology. Finally, we make the summary and conclusions in Sec. V.
\section{Model setup}\label{sec:model}
Typically speaking, there are six type of scalar LQs \cite{Dorsner:2016wpm}, which carry a conserved quantum number $F\equiv3B+L$. Here, the $B$ and $L$ are the baryon and lepton numbers. As to the VLQs, there are seven typical representations \cite{Aguilar-Saavedra:2013qpa}. In Tab. \ref{tab:LQ+VLQ:rep}, we list their representations and labels.
\begin{table}[!htb]
\begin{center}
\begin{tabular}{c|c|c}
\hline
\makecell{$SU(3)_C\times SU(2)_L\times U(1)_Y$ \\ representation} & label & $F$ \\
\hline
\rule{0pt}{10pt} $(\bar{3},3,1/3)$ & $S_3$ & $-2$ \\
\hline
$(3,2,7/6)$ & $R_2$ & 0 \\
\hline
\rule{0pt}{10pt} $(3,2,1/6)$ & $\widetilde{R}_2$ & 0 \\
\hline
\rule{0pt}{10pt} $(\bar{3},1,4/3)$ & $\widetilde{S}_1$ & $-2$ \\
\hline
\rule{0pt}{10pt} $(\bar{3},1,1/3)$ & $S_1$ & $-2$ \\
\hline
\rule{0pt}{10pt} $(\bar{3},1,-2/3)$ & $\bar{S}_1$ & $-2$ \\
\hline
\end{tabular}\qquad\qquad
\begin{tabular}{c|c}
\hline
\makecell{$SU(3)_C\times SU(2)_L\times U(1)_Y$ \\ representation} & label \\
\hline
$(3,1,2/3)$ & $T_{L,R}$ \\
\hline
$(3,1,-1/3)$ & $B_{L,R}$ \\
\hline
$(3,2,7/6)$ & $(X,T)_{L,R}$ \\
\hline
$(3,2,1/6)$ & $(T,B)_{L,R}$ \\
\hline
$(3,2,-5/6)$ & $(B,Y)_{L,R}$ \\
\hline
$(3,3,2/3)$ & $(X,T,B)_{L,R}$ \\
\hline
$(3,3,-1/3)$ & $(T,B,Y)_{L,R}$ \\
\hline
\end{tabular}
\caption{The scalar LQ (left) and VLQ (right) representations.} \label{tab:LQ+VLQ:rep}
\end{center}
\end{table}

For the six type of scalar LQs and seven type of VLQs, there can be totally 42 combinations, which are named as ``$\mr{LQ}+\mr{VLQ}$" for convenience. While, only 17 of them can lead to the chiral enhancements. In Tab. \ref{tab:LQ+VLQ:chiral}, we list these models that feature the chiral enhancements. The contributons in the four models $R_2+B_{L,R}/(B,Y)_{L,R}$ and $S_1+B_{L,R}/(B,Y)_{L,R}$ are almost the same as those in the minimal $R_2$ and $S_1$ models. There are nine models $R_2+T_{L,R}/(X,T)_{L,R}/(T,B)_{L,R}/(T,B,Y)_{L,R}$ and $S_1+T_{L,R}/(X,T)_{L,R}/(T,B)_{L,R}/(X,T,B)_{L,R}/(T,B,Y)_{L,R}$, which produce the top and top partner chiral enhancements. For the two models $R_2/S_3+(X,T,B)_{L,R}$, there are top, top partner, bottom, and bottom partner chiral enhancements at the same time. The models including $T$ quark have already been investigated in our paper \cite{He:2021yck}. Here, we will study the pure bottom partner chirally enhanced models $\widetilde{R}_2/\widetilde{S}_1+(B,Y)_{L,R}$.
\begin{table}[!htb]
\begin{center}
\begin{tabular}{c|c}
\hline
Model & chiral enhancement \\
\hline
$R_2$ & $m_t/m_{\mu}$ \\
\hline
$S_1$ & $m_t/m_{\mu}$ \\
\hline
$R_2+B_{L,R}/(B,Y)_{L,R}$ & $m_t/m_{\mu}$ \\
\hline
$S_1+B_{L,R}/(B,Y)_{L,R}$ & $m_t/m_{\mu}$ \\
\hline
$R_2+T_{L,R}/(X,T)_{L,R}/(T,B)_{L,R}/(T,B,Y)_{L,R}$ & $m_t/m_{\mu},m_T/m_{\mu}$ \\
\hline
$S_1+T_{L,R}/(X,T)_{L,R}/(T,B)_{L,R}/(X,T,B)_{L,R}/(T,B,Y)_{L,R}$ & $m_t/m_{\mu},m_T/m_{\mu}$ \\
\hline
$R_2+(X,T,B)_{L,R}$ & $m_t/m_{\mu},m_T/m_{\mu},m_b/m_{\mu},m_B/m_{\mu}$ \\
\hline
$S_3+(X,T,B)_{L,R}$ & $m_t/m_{\mu},m_T/m_{\mu},m_b/m_{\mu},m_B/m_{\mu}$ \\
\hline
\rule{0pt}{10pt} $\mathbf{\widetilde{R}_2+(B,Y)_{L,R}}$ & $m_b/m_{\mu},m_B/m_{\mu}$ \\
\hline
\rule{0pt}{10pt} $\mathbf{\widetilde{S}_1+(B,Y)_{L,R}}$ & $m_b/m_{\mu},m_B/m_{\mu}$  \\
\hline
\end{tabular}
\caption{Chiral enhancements in the minimal LQ and LQ+VLQ models.} \label{tab:LQ+VLQ:chiral}
\end{center}
\end{table}

\subsection{The VLQ Yukawa interactions with Higgs}
Let us start with the $(B,Y)_{L,R}$ related Higgs Yukawa interactions. In the gauge eigenstates, there are two interactions $\overline{Q_L}^id_R^j\phi$ and $(\overline{B_L},\overline{Y_L})d_R^j\widetilde{\phi}$, and the mass term $-M_B(\bar{B}B+\bar{Y}Y)$. Here, we define the SM Higgs doublet $\widetilde{\phi}\equiv i\sigma^2\phi^{\ast}$ with $\sigma^a(a=1,2,3)$ to be the Pauli matrices. The $Q_L^i$ and $d_R^i$ ($i=1,2,3$) represent the SM quark fields. We can parametrize $\phi$ as $[0,(v+h)/{\sqrt{2}}]^T$ in the unitary gauge. After the electroweak symmetry breaking (EWSB), there are mixings between $d^i$ and $B$. For simplicity, we only consider mixing between the third generation and $B$ quark. Thus, we can perform the following transformations to rotate $b$ and $B$ quarks into mass eigenstates:
\begin{align}\label{eqn:quarkbB:rotation}
\left[\begin{array}{c}b_L\\B_L\end{array}\right]\rightarrow
	\left[\begin{array}{cc}c_L^b&s_L^b\\-s_L^b&c_L^b\end{array}\right]
	\left[\begin{array}{c}b_L\\B_L\end{array}\right],\quad
\left[\begin{array}{c}b_R\\B_R\end{array}\right]\rightarrow
	\left[\begin{array}{cc}c_R^b&s_R^b\\-s_R^b&c_R^b\end{array}\right]
	\left[\begin{array}{c}b_R\\B_R\end{array}\right].
\end{align}
In the above, the $s_{L,R}^b$ and $c_{L,R}^b$ are abbreviations of $\sin\theta_{L,R}^b$ and $\cos\theta_{L,R}^b$. In fact, the $\theta_L^b$ can be correlated with $\theta_R^b$ through the relation $\tan\theta_L^b=m_b\tan\theta_R^b/m_B$ \cite{Aguilar-Saavedra:2013qpa}. Here, the $m_b$ and $m_B$ label the physical $b$ and $B$ quark masses. Besides, the mass of $Y$ quark is $m_Y=M_B=\sqrt{m_B^2(c_R^b)^2+m_b^2(s_R^b)^2}$. Then, we can choose $m_B$ and $\theta_R^b$ as the new input parameters. After the transformations in Eq. \eqref{eqn:quarkbB:rotation}, we obtain the following mass eigenstate Higgs Yukawa interactions:
\begin{align}
&\mathcal{L}_H^{Yukawa}\supset-\frac{m_b}{v}(c_R^b)^2h\bar{b}b-\frac{m_B}{v}(s_R^b)^2h\bar{B}B-\frac{m_b}{v}s_R^bc_R^bh(\bar{b}_LB_R+\bar{B}_Rb_L)-\frac{m_B}{v}s_R^bc_R^bh(\bar{B}_Lb_R+\bar{b}_RB_L).
\end{align}
Note that the $Y$ quark does not interact with Higgs at tree level.

\subsection{The VLQ gauge interactions}
Now, let us label the $SU(2)_L$ and $U_Y(1)$ gauge fields as $W_{\mu}^a$ and $B_{\mu}$. Then, the electroweak covariant derivative $D_{\mu}$ is defined as $\partial_{\mu}-igW_{\mu}^a\sigma^a/2-ig^{\prime}Y_qB_{\mu}$ for doublet and $\partial_{\mu}-ig^{\prime}Y_qB_{\mu}$ for singlet, in which the $Y_q$ is $U_Y(1)$ charge of the quark field acted by $D_{\mu}$. Thus, the related gauge interactions can be written as $\overline{Q_L}^iiD_{\mu}\gamma^{\mu}Q_L^i+\overline{d_R}^iiD_{\mu}\gamma^{\mu}d_R^i+(\overline{B},\overline{Y})iD_{\mu}\gamma^{\mu}(B,Y)^T$. After the EWSB, the $W$ gauge interactions can be written as
\begin{align}
\mc{L}\supset \frac{g}{\sqrt{2}}W_{\mu}^+(\overline{t_L}\gamma^{\mu}b_L+\bar{B}\gamma^{\mu}Y)+\mathrm{h.c.}.
\end{align}
The $Z$ gauge interactions can be written as
\begin{align}
&\mc{L}\supset\frac{g}{c_W}[(-\frac{1}{2}+\frac{1}{3}s_W^2)\overline{b_L}\gamma^{\mu}b_L+\frac{1}{3}s_W^2\overline{b_R}\gamma^{\mu}b_R+(\frac{1}{2}+\frac{1}{3}s_W^2)\bar{B}\gamma^{\mu}B+(-\frac{1}{2}+\frac{4}{3}s_W^2)\bar{Y}\gamma^{\mu}Y]Z_{\mu}.
\end{align}
After the rotations in Eq. \eqref{eqn:quarkbB:rotation}, we have the mass eigenstate $W$ gauge interactions:
\begin{align}
&\mc{L}_{BY}^{gauge}\supset\frac{g}{\sqrt{2}}W_{\mu}^+[c_L^b\overline{t_L}\gamma^{\mu}b_L+s_L^b\overline{t_L}\gamma^{\mu}B_L+c_L^b\overline{B_L}\gamma^{\mu}Y_L-s_L^b\overline{b_L}\gamma^{\mu}Y_L+c_R^b\overline{B_R}\gamma^{\mu}Y_R-s_R^b\overline{b_R}\gamma^{\mu}Y_R]+\mathrm{h.c.}.
\end{align}
We also have the mass eigenstate $Z$ gauge interactions:
\begin{align}\label{eqn:VLQ:Z}
&\mc{L}_{BY}^{gauge}\supset\frac{g}{c_W}Z_{\mu}[\frac{(c_L^b)^2-(s_L^b)^2}{2}(\overline{B_L}\gamma^{\mu}B_L-\overline{b_L}\gamma^{\mu}b_L)-s_L^bc_L^b(\overline{b_L}\gamma^{\mu}B_L+\overline{B_L}\gamma^{\mu}b_L)+\frac{(s_R^b)^2}{2}\overline{b_R}\gamma^{\mu}b_R\nonumber\\
&+\frac{(c_R^b)^2}{2}\overline{B_R}\gamma^{\mu}B_R-\frac{s_R^bc_R^b}{2}(\overline{b_R}\gamma^{\mu}B_R+\overline{B_R}\gamma^{\mu}b_R)+\frac{s_W^2}{3}(\bar{b}\gamma^{\mu}b+\bar{B}\gamma^{\mu}B)+(-\frac{1}{2}+\frac{4s_W^2}{3})\bar{Y}\gamma^{\mu}Y].
\end{align}

\subsection{The VLQ Yukawa interactions with LQ}
Let us denote the SM lepton fields as $L_L^i$ and $e_R^i$. The $\widetilde{R}_2$ can be parametrized as $[\widetilde{R}_2^{2/3},\widetilde{R}_2^{-1/3}]^T$, where the superscript labels the electric charge. Then, the $\widetilde{R}_2$ and $\widetilde{S}_1$ can induce the following $F=0$ and $F=2$ type gauge eigenstate LQ Yukawa interactions:
\begin{align}
\mc{L}_{\widetilde{R}_2+(B,Y)_{L,R}}\supset x_i\overline{e_R^i}(\widetilde{R}_2)^{\dag}\left(\begin{array}{c}B_L\\Y_L\end{array}\right)+y_{ij}\overline{L_L^i}\epsilon(\widetilde{R}_2)^\ast d_R^j+\mathrm{h.c.},
\end{align}
and
\begin{align}
\mc{L}_{\widetilde{S}_1+(B,Y)_{L,R}}\supset x_{ij}\overline{e_R^i}(d_R^j)^C(\widetilde{S}_1)^{\ast}+y_i\overline{L_L^i}\epsilon\left(\begin{array}{c}B_L\\Y_L\end{array}\right)^C(\widetilde{S}_1)^\ast+\mathrm{h.c.}.
\end{align}
After the EWSB, they can be parametrized as
\begin{align}
&\mc{L}_{\widetilde{R}_2+(B,Y)_{L,R}}\supset y_L^{\widetilde{R}_2\mu B}\bar{\mu}~\omega_-B(\widetilde{R}_2^{2/3})^\ast+y_R^{\widetilde{R}_2\mu b}\bar{\mu}~\omega_+b(\widetilde{R}_2^{2/3})^\ast\nonumber\\
&+y_L^{\widetilde{R}_2\mu B}\bar{\mu}~\omega_-Y(\widetilde{R}_2^{-1/3})^\ast-y_R^{\widetilde{R}_2\mu b}\overline{\nu_L}~\omega_+b(\widetilde{R}_2^{-1/3})^\ast+\mathrm{h.c.},
\end{align}
and
\begin{align}
\mc{L}_{\widetilde{S}_1+(B,Y)_{L,R}}\supset y_L^{\widetilde{S}_1\mu b}\bar{\mu}~\omega_-b^C(\widetilde{S}_1)^\ast+y_R^{\widetilde{S}_1\mu B}\bar{\mu}~\omega_+B^C(\widetilde{S}_1)^\ast-y_R^{\widetilde{S}_1\mu B}\overline{\nu_L}~\omega_+Y^C(\widetilde{S}_1)^\ast+\mathrm{h.c.}.
\end{align}
In the above, we define the chiral operators $\omega_{\pm}$ as $(1\pm\gamma^5)/2$. After the rotations in Eq. \eqref{eqn:quarkbB:rotation}, we have the mass eigenstate interactions:
\begin{align}
&\mc{L}_{\widetilde{R}_2+(B,Y)_{L,R}}\supset \bar{\mu}(-y_L^{\widetilde{R}_2\mu B}s_L^b\omega_-+y_R^{\widetilde{R}_2\mu b}c_R^b\omega_+)b(\widetilde{R}_2^{2/3})^\ast+\bar{\mu}(y_L^{\widetilde{R}_2\mu B}c_L^b\omega_-+y_R^{\widetilde{R}_2\mu b}s_R^b\omega_+)B(\widetilde{R}_2^{2/3})^\ast\nonumber\\
&+y_L^{\widetilde{R}_2\mu B}\bar{\mu}~\omega_-Y(\widetilde{R}_2^{-1/3})^\ast-y_R^{\widetilde{R}_2\mu b}\overline{\nu_L}\omega_+(c_R^bb+s_R^bB)(\widetilde{R}_2^{-1/3})^\ast+\mathrm{h.c.},
\end{align}
and
\begin{align}
&\mc{L}_{\widetilde{S}_1+(B,Y)_{L,R}}\supset \bar{\mu}(y_L^{\widetilde{S}_1\mu b}c_R^b\omega_--y_R^{\widetilde{S}_1\mu B}s_L^b\omega_+)b^C(\widetilde{S}_1)^\ast+\bar{\mu}(y_L^{\widetilde{S}_1\mu b}s_R^b\omega_-+y_R^{\widetilde{S}_1\mu B}c_L^b\omega_+)B^C(\widetilde{S}_1)^\ast\nonumber\\
&-y_R^{\widetilde{S}_1\mu B}\overline{\nu_L}~\omega_+Y^C(\widetilde{S}_1)^\ast+\mathrm{h.c.}.
\end{align}
\section{Contributions to the $(g-2)_{\mu}$}\label{sec:g-2}
\subsection{Analytic results of the contributions}
For the $\mr{LQ}\mu q$ interaction, there are quark-photon and LQ-photon vertex mediated contributions to the $(g-2)_{\mu}$, which can be described by the functions $f^q(x)$ and $f^S(x)$. Then, we use the functions $f_{LL}^{q,S}(x)$ and $f_{LR}^{q,S}(x)$ to label the parts without and with chiral enhancements. Starting from the $f_{LL}^{q,S}(x)$ and $f_{LR}^{q,S}(x)$ given in our paper \cite{He:2021yck}, let us define the following integrals:
\begin{align}\label{eqn:LQ:R2S1b}
&f_{LL}^{\widetilde{R}_2\mu Y}(x)\equiv 4f_{LL}^q(x)-f_{LL}^S(x)=\frac{3+2x-7x^2+2x^3+2x(4-x)\log x}{4(1-x)^4},\nonumber\\
&f_{LL}^{\widetilde{R}_2\mu b}(x)\equiv f_{LL}^q(x)+2f_{LL}^S(x)=\frac{x[5-4x-x^2+(2+4x)\log x]}{4(1-x)^4},\nonumber\\
&f_{LR}^{\widetilde{R}_2\mu b}(x)\equiv f_{LR}^q(x)+2f_{LR}^S(x)=-\frac{5-4x-x^2+(2+4x)\log x}{4(1-x)^3},\nonumber\\
&f_{LL}^{\widetilde{S}_1\mu b}(x)\equiv-f_{LL}^q(x)+4f_{LL}^S(x)=-\frac{2-7x+2x^2+3x^3+2x(1-4x)\log x}{4(1-x)^4},\nonumber\\
&f_{LR}^{\widetilde{S}_1\mu b}(x)\equiv-f_{LR}^q(x)+4f_{LR}^S(x)=-\frac{1+4x-5x^2-(2-8x)\log x}{4(1-x)^3}.
\end{align}
For the $\widetilde{R}_2+(B,Y)_{L,R}$ model, there are $b$, $B$, and $Y$ quark contributions to the $(g-2)_{\mu}$. The complete expression is calculated as
\begin{align}
&\Delta a_{\mu}^{\widetilde{R}_2+BY}=\frac{m_{\mu}^2}{8\pi^2}\Bigg\{\frac{|y_L^{\widetilde{R}_2\mu B}|^2}{m_{\widetilde{R}_2^{-1/3}}^2}f_{LL}^{\widetilde{R}_2\mu Y}(\frac{m_Y^2}{m_{\widetilde{R}_2^{-1/3}}^2})\nonumber\\
&+\frac{|y_L^{\widetilde{R}_2\mu B}|^2(s_L^b)^2+|y_R^{\widetilde{R}_2\mu b}|^2(c_R^b)^2}{m_{\widetilde{R}_2^{2/3}}^2}f_{LL}^{\widetilde{R}_2\mu b}(\frac{m_b^2}{m_{\widetilde{R}_2^{2/3}}^2})-\frac{2m_b}{m_{\mu}}\frac{s_L^bc_R^b}{m_{\widetilde{R}_2^{2/3}}^2}\mr{Re}[y_L^{\widetilde{R}_2\mu B}(y_R^{\widetilde{R}_2\mu b})^\ast]f_{LR}^{\widetilde{R}_2\mu b}(\frac{m_b^2}{m_{\widetilde{R}_2^{2/3}}^2})\nonumber\\
&+\frac{|y_L^{\widetilde{R}_2\mu B}|^2(c_L^b)^2+|y_R^{\widetilde{R}_2\mu b}|^2(s_R^b)^2}{m_{\widetilde{R}_2^{2/3}}^2}f_{LL}^{\widetilde{R}_2\mu b}(\frac{m_B^2}{m_{\widetilde{R}_2^{2/3}}^2})+\frac{2m_B}{m_{\mu}}\frac{c_L^bs_R^b}{m_{\widetilde{R}_2^{2/3}}^2}\mr{Re}[y_L^{\widetilde{R}_2\mu B}(y_R^{\widetilde{R}_2\mu b})^\ast]f_{LR}^{\widetilde{R}_2\mu b}(\frac{m_B^2}{m_{\widetilde{R}_2^{2/3}}^2})\Bigg\}.
\end{align}
At tree level, we have $m_{\widetilde{R}_2^{2/3}}=m_{\widetilde{R}_2^{-1/3}}\equiv m_{\widetilde{R}_2}$. Compared to the bottom partner chirally enhanced contribution (namely, the $f_{LR}^{\widetilde{R}_2\mu b}(\frac{m_B^2}{m_{\widetilde{R}_2^{2/3}}^2})$ related term), the non-chirally enhanced parts are suppressed by the factor $m_{\mu}/(m_Bs_R^b)\sim1/(10^4s_R^b)$ and the bottom quark chirally enhanced part is suppressed by the factor $(m_bs_L^b)/(m_Bs_R^b)\sim(m_b^2/m_B^2)$. For the interesting values of $s_R^b$ at $\mc{O}(0.01\sim0.1)$, the $\Delta a_{\mu}^{\widetilde{R}_2+BY}$ is dominated by the bottom partner chirally enhanced contribution. Then, the above expression can be approximated as
\begin{align}\label{eqn:g-2:R2app}
&\Delta a_{\mu}^{\widetilde{R}_2+BY}\approx\frac{m_{\mu}m_B}{4\pi^2m_{\widetilde{R}_2}^2}s_R^b\mr{Re}[y_L^{\widetilde{R}_2\mu B}(y_R^{\widetilde{R}_2\mu b})^\ast]f_{LR}^{\widetilde{R}_2\mu b}(\frac{m_B^2}{m_{\widetilde{R}_2}^2}).
\end{align}

For the $\widetilde{S}_1+(B,Y)_{L,R}$ model, there are $b$ and $B$ quark contributions to the $(g-2)_{\mu}$. The complete expression is calculated as
\begin{align}
&\Delta a_{\mu}^{\widetilde{S}_1+BY}=\frac{m_{\mu}^2}{8\pi^2}\Bigg\{\nonumber\\
&\frac{|y_R^{\widetilde{S}_1\mu B}|^2(s_L^b)^2+|y_L^{\widetilde{S}_1\mu b}|^2(c_R^b)^2}{m_{\widetilde{S}_1}^2}f_{LL}^{\widetilde{S}_1\mu b}(\frac{m_b^2}{m_{\widetilde{S}_1}^2})-\frac{2m_b}{m_{\mu}}\frac{s_L^bc_R^b}{m_{\widetilde{S}_1}^2}\mr{Re}[y_R^{\widetilde{S}_1\mu B}(y_L^{\widetilde{S}_1\mu b})^\ast]f_{LR}^{\widetilde{S}_1\mu b}(\frac{m_b^2}{m_{\widetilde{S}_1}^2})\nonumber\\
&+\frac{|y_R^{\widetilde{S}_1\mu B}|^2(c_L^b)^2+|y_L^{\widetilde{S}_1\mu b}|^2(s_R^b)^2}{m_{\widetilde{S}_1}^2}f_{LL}^{\widetilde{S}_1\mu b}(\frac{m_B^2}{m_{\widetilde{S}_1}^2})+\frac{2m_B}{m_{\mu}}\frac{c_L^bs_R^b}{m_{\widetilde{S}_1}^2}\mr{Re}[y_R^{\widetilde{S}_1\mu B}(y_L^{\widetilde{S}_1\mu b})^\ast]f_{LR}^{\widetilde{S}_1\mu b}(\frac{m_B^2}{m_{\widetilde{S}_1}^2})\Bigg\}.
\end{align}
Similarly, it can be approximated as
\begin{align}\label{eqn:g-2:S1app}
&\Delta a_{\mu}^{\widetilde{S}_1+BY}\approx\frac{m_{\mu}m_B}{4\pi^2m_{\widetilde{S}_1}^2}s_R^b\mr{Re}[y_R^{\widetilde{S}_1\mu B}(y_L^{\widetilde{S}_1\mu b})^\ast]f_{LR}^{\widetilde{S}_1\mu b}(\frac{m_B^2}{m_{\widetilde{S}_1}^2}).
\end{align}

\subsection{Numerical analysis}
The input parameters are chosen as $m_\mu=105.66\mathrm{MeV}$, $m_b=4.2\mr{GeV}$, $m_t=172.5\mr{GeV}$, $G_F=1.1664\times10^{-5}\mr{GeV}^{-2}$, $m_W=80.377\mr{GeV}$, $m_Z=91.1876\mr{GeV}$, $m_h=125.25\mr{GeV}$ \cite{Workman:2022ynf}. Besides, the $v,g,\theta_W$ are defined by $G_F=1/(\sqrt{2}v^2),g=2m_W/v,\cos\theta_W\equiv c_W=m_W/m_Z$. There are also new parameters $m_B$, $m_{\mr{LQ}}$, $\theta_R^b$, and the LQ Yukawa couplings $y_{L,R}^{\mr{LQ}\mu q}$. The VLQ mass can be constrained from the direct search, which is required to be above $1.5\mr{TeV}$ \cite{CMS:2020ttz, CMS:2017fpk, ATLAS:2018ziw, ATLAS:2018dyh}. The mixing angle is mainly bounded by the electro-weak precision observables (EWPOs). The VLQ contributions to $T$ parameter are suppressed by the factor $(s_R^b)^4$ or $m_b^2(s_R^b)^2/m_B^2$ \cite{Chen:2017hak, Cao:2022mif}, thus it leads to less constrained $\theta_R^b$. The weak isospin third component of $B_R$ is positive, then the mixing with bottom quark enhances the right-handed $Zbb$ coupling. As a result, the $A_{FB}^b$ deviation \cite{ALEPH:2005ab, Baak:2014ora} can be compensated, which also leads to looser constraints on $\theta_R^b$. Conservatively speaking, we can choose the mixing angle $s_R^b$ to be less than $0.1$ \cite{Aguilar-Saavedra:2013qpa}. The LQ mass can also be constrained from the direct search, which is required to be above $1.7\mr{TeV}$ assuming $\mr{Br}(\mr{LQ}\rightarrow b\mu)=1$ \cite{CMS:2018lab, ATLAS:2020dsk}.

We can choose benchmark points of $m_B,m_{\mr{LQ}},s_R^b$ to constrain the LQ Yukawa couplings. Here, we consider two scenarios $m_{\mr{LQ}}>m_B$ and $m_{\mr{LQ}}<m_B$. For the scenario $m_{\mr{LQ}}>m_B$, we adopt the mass parameters to be $m_B=1.5\mr{TeV}$ and $m_{\mr{LQ}}=2\mr{TeV}$. For the scenario $m_{\mr{LQ}}<m_B$, we adopt the mass parameters to be $m_B=2.5\mr{TeV}$ and $m_{\mr{LQ}}=2\mr{TeV}$. In Tab. \ref{tab:LQ+BY:g-2num}, we give the approximate numerical expressions of the $\Delta a_{\mu}$ in the $\widetilde{R}_2/\widetilde{S}_1+(B,Y)_{L,R}$ models. Besides, we also show the allowed ranges for $s_R^b=0.1$ and $s_R^b=0.05$.
\begin{table}[!htb]
\begin{center}
\begin{tabular}{c|c|c|c|c|c}
\hline
\rule{0pt}{12pt}\multirow{2}{*}{Model} & \multirow{2}{*}{$(m_B,m_{\mr{LQ}})/\mr{TeV}$} & \multirow{2}{*}{$\Delta a_{\mu}\times10^7$} & \multirow{2}{*}{$s_R^b$} & \multicolumn{2}{c}{$\mr{Re}[y_L^{\widetilde{R}_2\mu B}(y_R^{\widetilde{R}_2\mu b})^\ast]$ or $\mr{Re}[y_R^{\widetilde{S}_1\mu B}(y_L^{\widetilde{S}_1\mu b})^\ast]$}\\ \cline{5-6}
& & & & $1\sigma$ & $2\sigma$\\
\hline
\multirow{4}{*}{$\widetilde{R}_2+(B,Y)_{L,R}$} & \multirow{2}{*}{$(1.5,2)$} & \multirow{2}{*}{$0.35s_R^b\mr{Re}[y_L^{\widetilde{R}_2\mu B}(y_R^{\widetilde{R}_2\mu b})^\ast]$} & 0.1 & $(0.55,0.88)$ & $(0.38,1.05)$\\
\cline{4-6}
& & & 0.05 & $(1.1,1.77)$ & $(0.76,2.11)$\\ \cline{2-6}

& \multirow{2}{*}{$(2.5,2)$} & \multirow{2}{*}{$-0.224s_R^b\mr{Re}[y_L^{\widetilde{R}_2\mu B}(y_R^{\widetilde{R}_2\mu b})^\ast]$} & 0.1 & $(-1.38,-0.86)$ & $(-1.65,-0.59)$\\ \cline{4-6}
& & & 0.05 & $(-2.76,-1.71)$ & $(-3.29,-1.19)$\\
\hline\hline
\multirow{4}{*}{$\widetilde{S}_1+(B,Y)_{L,R}$} & \multirow{2}{*}{$(1.5,2)$} & \multirow{2}{*}{$-6.88s_R^b\mr{Re}[y_R^{\widetilde{S}_1\mu B}(y_L^{\widetilde{S}_1\mu b})^\ast]$} & 0.1 & $(-0.045,-0.028)$ & $(-0.054,-0.019)$\\
\cline{4-6}
& & & 0.05 & $(-0.09,-0.056)$ & $(-0.11,-0.039)$\\ \cline{2-6}

& \multirow{2}{*}{$(2.5,2)$} & \multirow{2}{*}{$-6.37s_R^b\mr{Re}[y_R^{\widetilde{S}_1\mu B}(y_L^{\widetilde{S}_1\mu b})^\ast]$} & 0.1 & $(-0.049,-0.03)$ & $(-0.058,-0.021)$\\ \cline{4-6}
& & & 0.05 & $(-0.097,-0.06)$ & $(-0.12,-0.042)$\\
\hline
\end{tabular}
\caption{In the third column, we show the leading order numerical expressions of the $\Delta a_{\mu}$. In the fifth and sixth columns, we show the ranges allowed by $(g-2)_{\mu}$ at $1\sigma$ and $2\sigma$ confidence levels (CLs).} \label{tab:LQ+BY:g-2num}
\end{center}
\end{table}
Of course, these behaviours can be understood from the Eqs. \eqref{eqn:g-2:R2app} and \eqref{eqn:g-2:S1app}. In the $\widetilde{R}_2+(B,Y)_{L,R}$ model, the $f_{LR}^{\widetilde{R}_2\mu b}(x)$ vanishes as $m_B= m_{\widetilde{R}_2}$, which causes the $|\mr{Re}[y_L^{\widetilde{R}_2\mu B}(y_R^{\widetilde{R}_2\mu b})^\ast]|$ to be under the stress of perturbative unitarity. If there is a large hierarchy between $m_B$ and $m_{\widetilde{R}_2}$, the allowed $|\mr{Re}[y_L^{\widetilde{R}_2\mu B}(y_R^{\widetilde{R}_2\mu b})^\ast]|$ can be smaller. Besides, the $\mr{Re}[y_L^{\widetilde{R}_2\mu B}(y_R^{\widetilde{R}_2\mu b})^\ast]$ should be positive (negative) when $m_B<m_{\widetilde{R}_2}$ ($m_B>m_{\widetilde{R}_2}$). In the $\widetilde{S}_1+(B,Y)_{L,R}$ model, the $f_{LR}^{\widetilde{S}_1\mu b}(x)$ is always negative, which requires $\mr{Re}[y_R^{\widetilde{S}_1\mu B}(y_L^{\widetilde{S}_1\mu b})^\ast]<0$.

\begin{figure}[!htb]
\begin{center}
\includegraphics[scale=0.39]{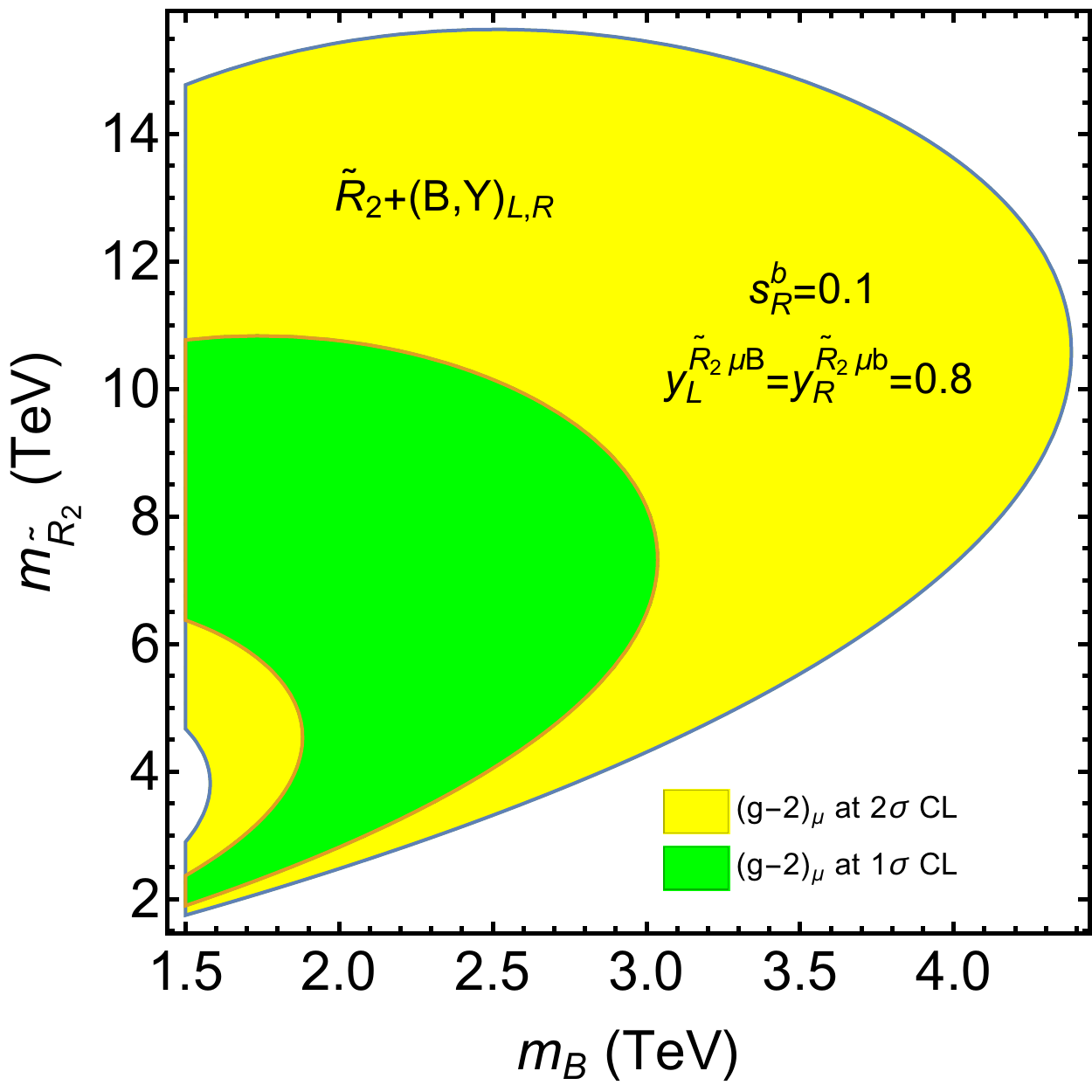}
\includegraphics[scale=0.39]{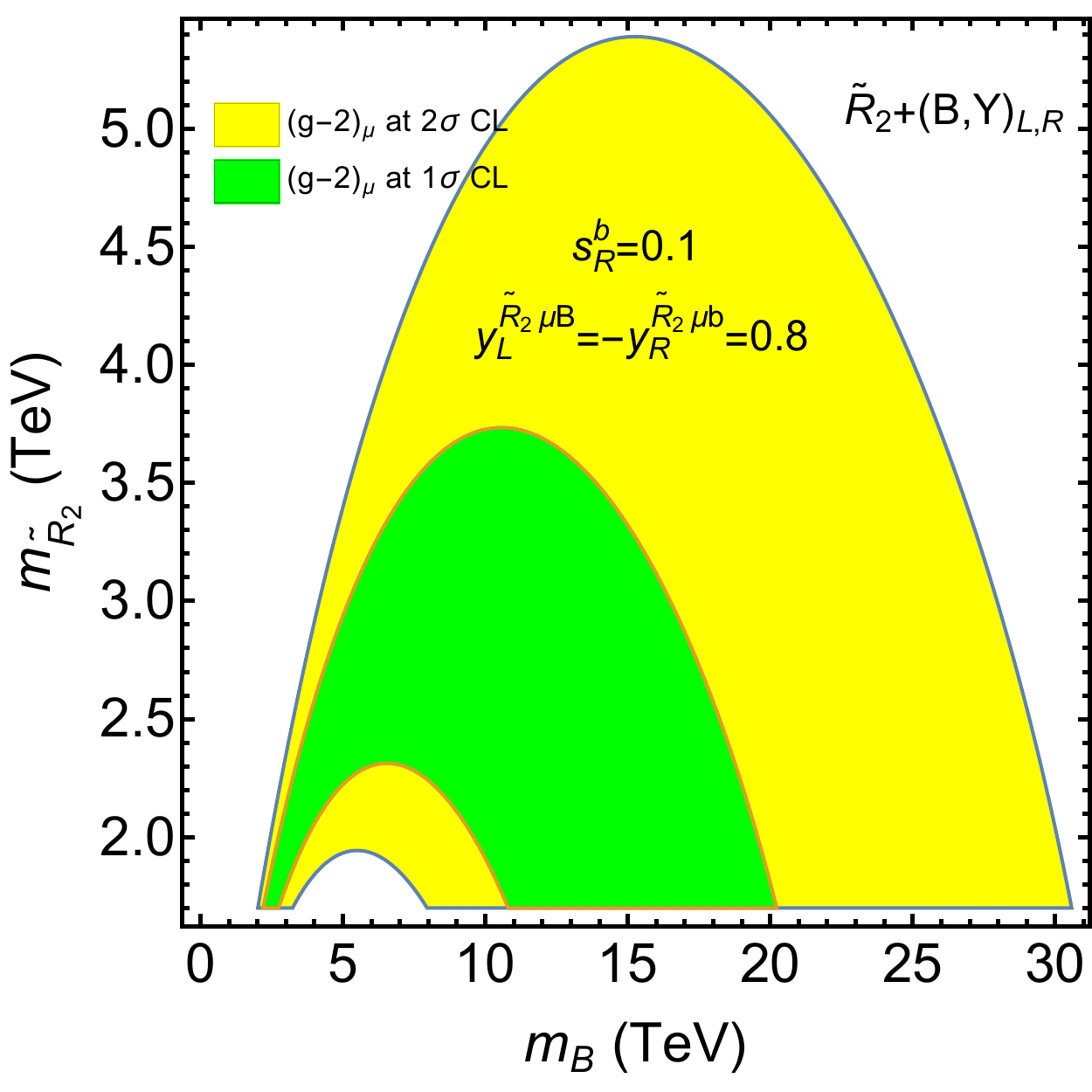}
\includegraphics[scale=0.39]{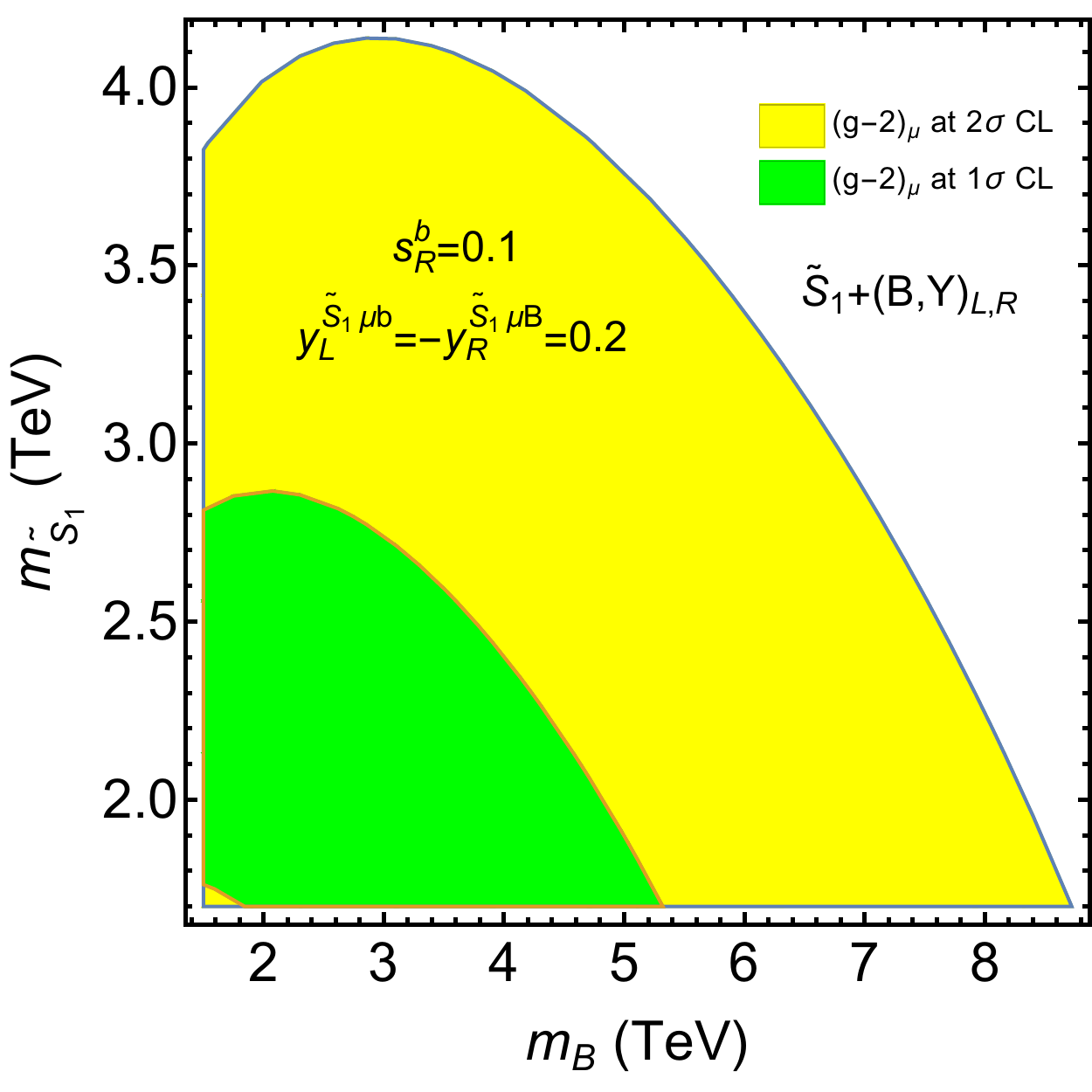}
\caption{The $(g-2)_{\mu}$ allowed regions at $1\sigma$ (green) and $2\sigma$ (yellow) CLs with $s_R^b=0.1$. The parameters are chosen as $y_L^{\widetilde{R}_2\mu B}=y_R^{\widetilde{R}_2\mu b}=0.8$ in the $\widetilde{R}_2+(B,Y)$ model (left), $y_L^{\widetilde{R}_2\mu B}=-y_R^{\widetilde{R}_2\mu b}=0.8$ in the $\widetilde{R}_2+(B,Y)$ model (middle), and $y_L^{\widetilde{S}_1\mu b}=-y_R^{\widetilde{S}_1\mu B}=0.2$ in the $\widetilde{S}_1+(B,Y)$ model (right).}\label{fig:LQ+BY:mBmLQ}
\end{center}
\end{figure}

We can also choose benchmark points of $s_R^b$ and LQ Yukawa couplings to constrain the $m_B$ and $m_{\mr{LQ}}$. In Fig. \ref{fig:LQ+BY:mBmLQ}, we show the $(g-2)_{\mu}$ allowed regions in the plane of $m_B-m_{\mr{LQ}}$. As we can see, the $m_B<m_{\widetilde{R}_2}$ and $m_B>m_{\widetilde{R}_2}$ are favoured in the left and middle plots, respectively. This can be understood from the asymptotic behaviours $f_{LR}^{\widetilde{R}_2\mu b}(x)\sim-\log(x)/2>0$ for $x\rightarrow0$ and $f_{LR}^{\widetilde{R}_2\mu b}(x)\sim-1/(4x)<0$ for $x\rightarrow\infty$. To produce positive $\Delta a_{\mu}$, the $m_B<m_{\widetilde{R}_2}$ and $m_B>m_{\widetilde{R}_2}$ are favoured for $\mr{Re}[y_L^{\widetilde{R}_2\mu B}(y_R^{\widetilde{R}_2\mu b})^\ast]>0$ and $\mr{Re}[y_L^{\widetilde{R}_2\mu B}(y_R^{\widetilde{R}_2\mu b})^\ast]<0$, respectively. The $f_{LR}^{\widetilde{S}_1\mu b}(x)$ has asymptotic behaviours $f_{LR}^{\widetilde{S}_1\mu b}(x)\sim\log(x)/2<0$ for $x\rightarrow0$ and $f_{LR}^{\widetilde{S}_1\mu b}(x)\sim-5/(4x)<0$ for $x\rightarrow\infty$. To produce positive $\Delta a_{\mu}$, the $\mr{Re}[y_R^{\widetilde{S}_1\mu B}(y_L^{\widetilde{S}_1\mu b})^\ast]<0$ is favoured. Furthermore, the allowed regions in the plane of $m_B-m_{\mr{LQ}}$ are sensitive to the choice of $y_{L,R}^{\mr{LQ}\mu q}$. Roughly speaking, the larger $|y_{L,R}^{\mr{LQ}\mu q}|$, the larger $m_B$ and $m_{\mr{LQ}}$.

\section{LQ and VLQ Phenomenology at hadron colliders}\label{sec:phenomenology}
In Tab. \ref{tab:LQ+BY:decay}, we list the main LQ and VLQ decay channels \footnote{The $\widetilde{R}_2^{-1/3}\rightarrow\nu_LB$ and $B\rightarrow\nu_L\widetilde{R}_2^{-1/3}$ decay channels are suppressed by the factor $(s_R^b)^2$. The $B\rightarrow tW^-$ decay channel is suppressed by the factor $m_b^2/m_B^2$.}. The decay formulae of LQ and VLQ are given in App. \ref{app:decay:LQ} and \ref{app:decay:VLQ}. For the scenario $m_{\mr{LQ}}>m_B$, there are new LQ decay channels. When searching for the LQ $\widetilde{R}_2^{2/3}$, we propose the $\mu j_bZ$ and $\mu j_bh$ signatures. When searching for the LQ $\widetilde{R}_2^{-1/3}$, we propose the $\mu j_bW$ signatures. When searching for the LQ $\widetilde{S}_1$, we propose the $\mu j_bZ$, $\mu j_bh$, $\slash\!\!\!\!{E_T}j_bW$ signatures. For the scenario $m_{\mr{LQ}}<m_B$, there are new VLQ decay channels. When searching for the VLQ $B$, we propose the $\mu^+\mu^-j_b$ signatures. When searching for the VLQ $Y$, we propose the $\mu~\slash\!\!\!\!{E_T}j_b$ signatures. It seems that such decay channels have not been searched by the experimental collaborations.
\begin{table}[!htb]
\begin{center}
\begin{tabular}{c|c|c|c|c}
\hline
Model & Scenario & LQ decay & VLQ decay & new signatures\\
\hline
\rule{0pt}{12pt}\multirow{4}{*}{$\widetilde{R}_2+(B,Y)_{L,R}$} & \multirow{2}{*}{$m_{\mr{LQ}}>m_B$} & $\widetilde{R}_2^{2/3}\rightarrow \mu^+b,\mu^+B$ & $B\rightarrow bZ,bh$ & $\widetilde{R}_2^{2/3}\rightarrow\mu j_bZ,\mu j_bh$ \\
\cline{3-5}
\rule{0pt}{12pt}& & $\widetilde{R}_2^{-1/3}\rightarrow \mu^+Y,\nu_Lb$ & $Y\rightarrow bW^-$ & $\widetilde{R}_2^{-1/3}\rightarrow\mu j_bW$ \\ \cline{2-5}

\rule{0pt}{12pt}& \multirow{2}{*}{$m_{\mr{LQ}}<m_B$} & $\widetilde{R}_2^{2/3}\rightarrow \mu^+b$ & $B\rightarrow bZ,bh,\mu^-\widetilde{R}_2^{2/3}$ & $B\rightarrow\mu^+\mu^-j_b$\\ \cline{3-5}
\rule{0pt}{12pt}& & $\widetilde{R}_2^{-1/3}\rightarrow \nu_Lb$ & $Y\rightarrow bW^-,\mu^-\widetilde{R}_2^{-1/3}$ & $Y\rightarrow\mu~\slash\!\!\!\!{E_T}j_b$\\
\hline\hline

\rule{0pt}{12pt}\multirow{4}{*}{$\widetilde{S}_1+(B,Y)_{L,R}$} & \multirow{2}{*}{$m_{\mr{LQ}}>m_B$} & \multirow{2}{*}{$\widetilde{S}_1\rightarrow \mu^+\bar{b},\mu^+\bar{B},\nu_L\bar{Y}$} & $B\rightarrow bZ,bh$ & \multirow{2}{*}{$\widetilde{S}_1\rightarrow\mu j_bZ$, $\mu j_bh$, $\slash\!\!\!\!{E_T}j_bW$}\\
\cline{4-4}
\rule{0pt}{12pt}& & & $Y\rightarrow bW^-$ & \\ \cline{2-5}

\rule{0pt}{12pt}& \multirow{2}{*}{$m_{\mr{LQ}}<m_B$} & \multirow{2}{*}{$\widetilde{S}_1\rightarrow \mu^+\bar{b}$} & $B\rightarrow bZ,bh,\mu^+(\widetilde{S}_1)^{\ast}$ & $B\rightarrow\mu^+\mu^-j_b$\\ \cline{4-5}
\rule{0pt}{12pt}& & & $Y\rightarrow bW^-,\nu_L(\widetilde{S}_1)^{\ast}$ & $Y\rightarrow\mu~\slash\!\!\!\!{E_T}j_b$\\
\hline
\end{tabular}
\caption{In the third column, we show the main LQ decay channels. In the fourth column, we show the main VLQ decay channels. In the fifth column, we show the new LQ or VLQ signatures.} \label{tab:LQ+BY:decay}
\end{center}
\end{table}

\begin{figure}[!htb]
\begin{center}
\includegraphics[scale=0.43]{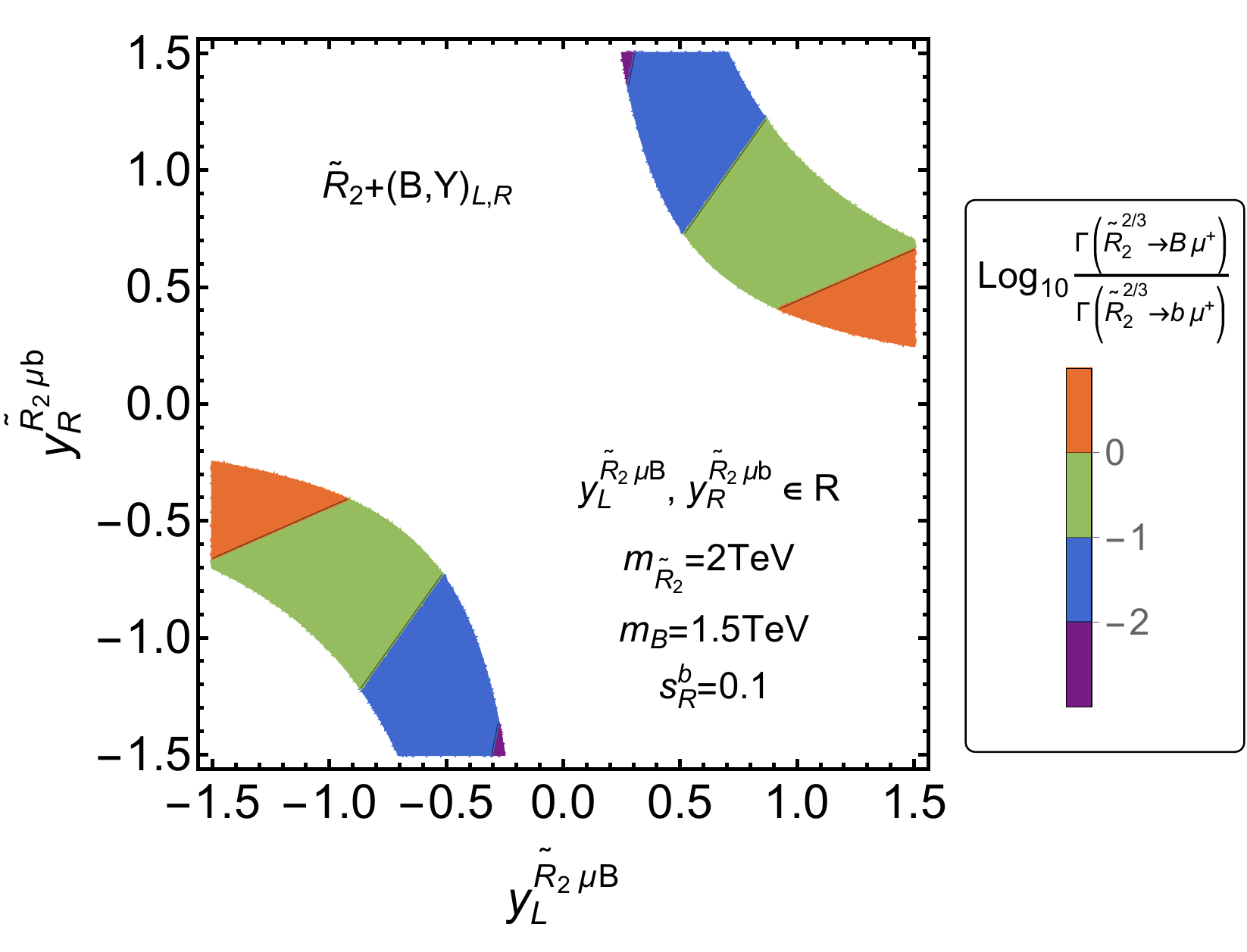}\qquad
\includegraphics[scale=0.43]{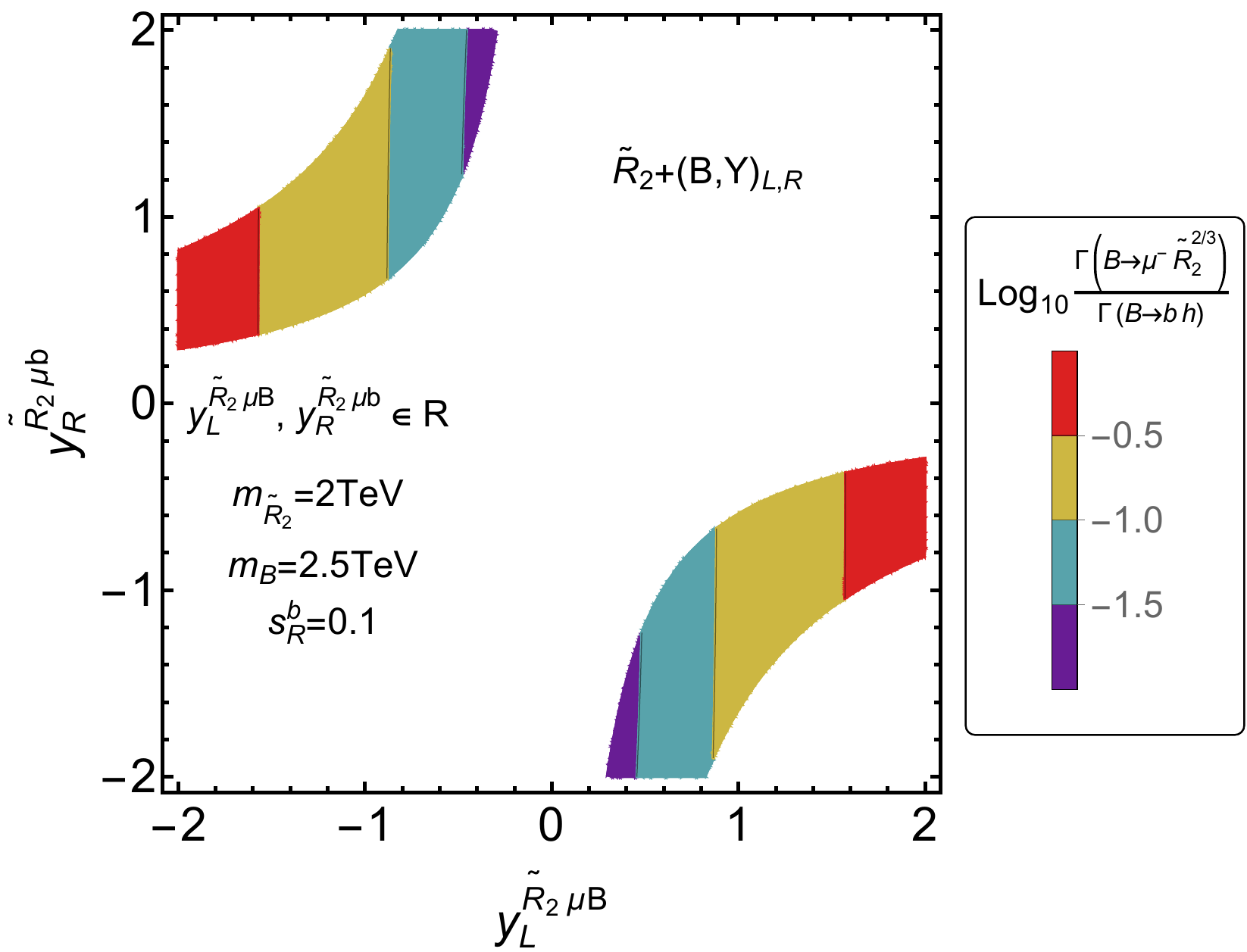}\\
\includegraphics[scale=0.43]{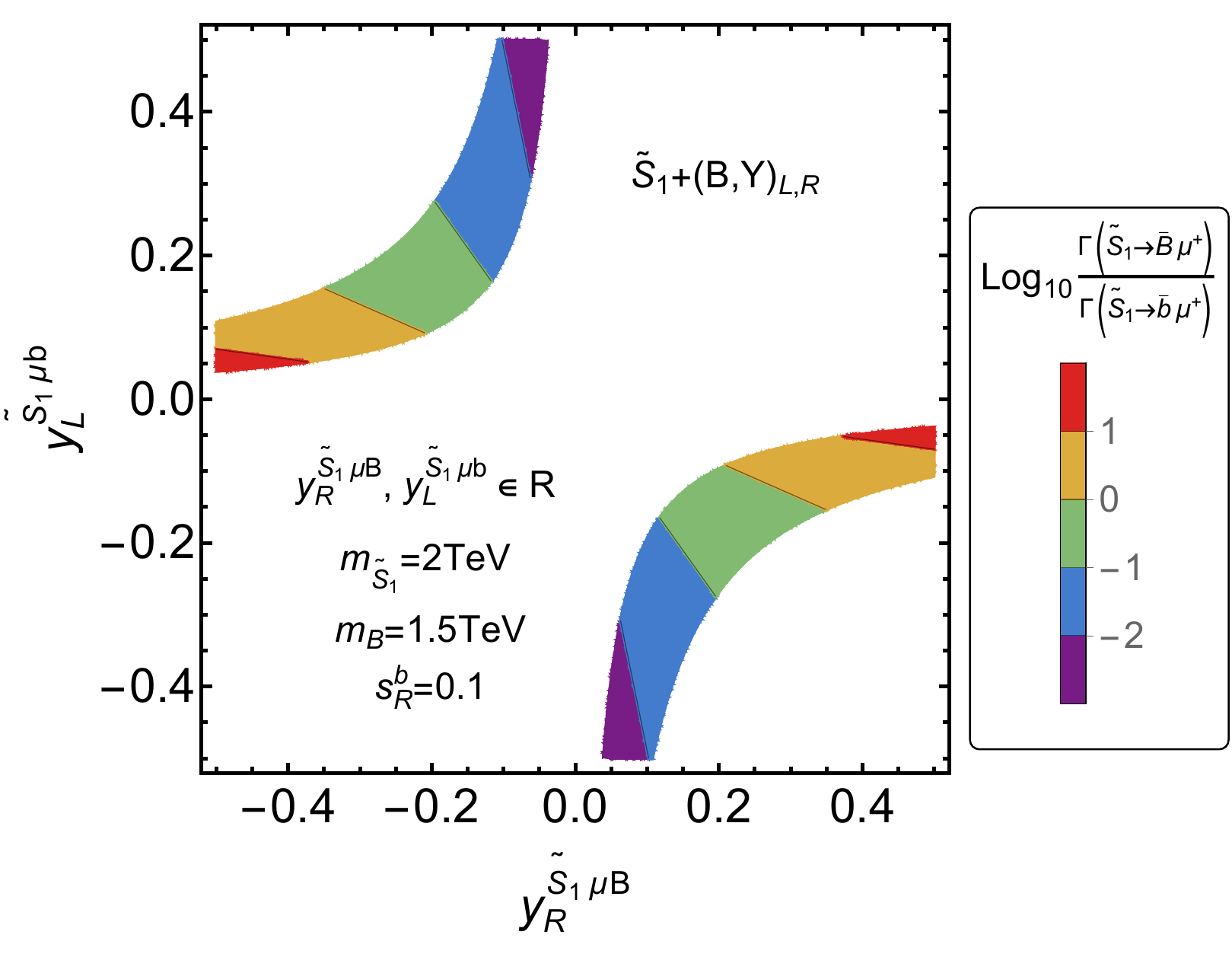}\qquad
\includegraphics[scale=0.43]{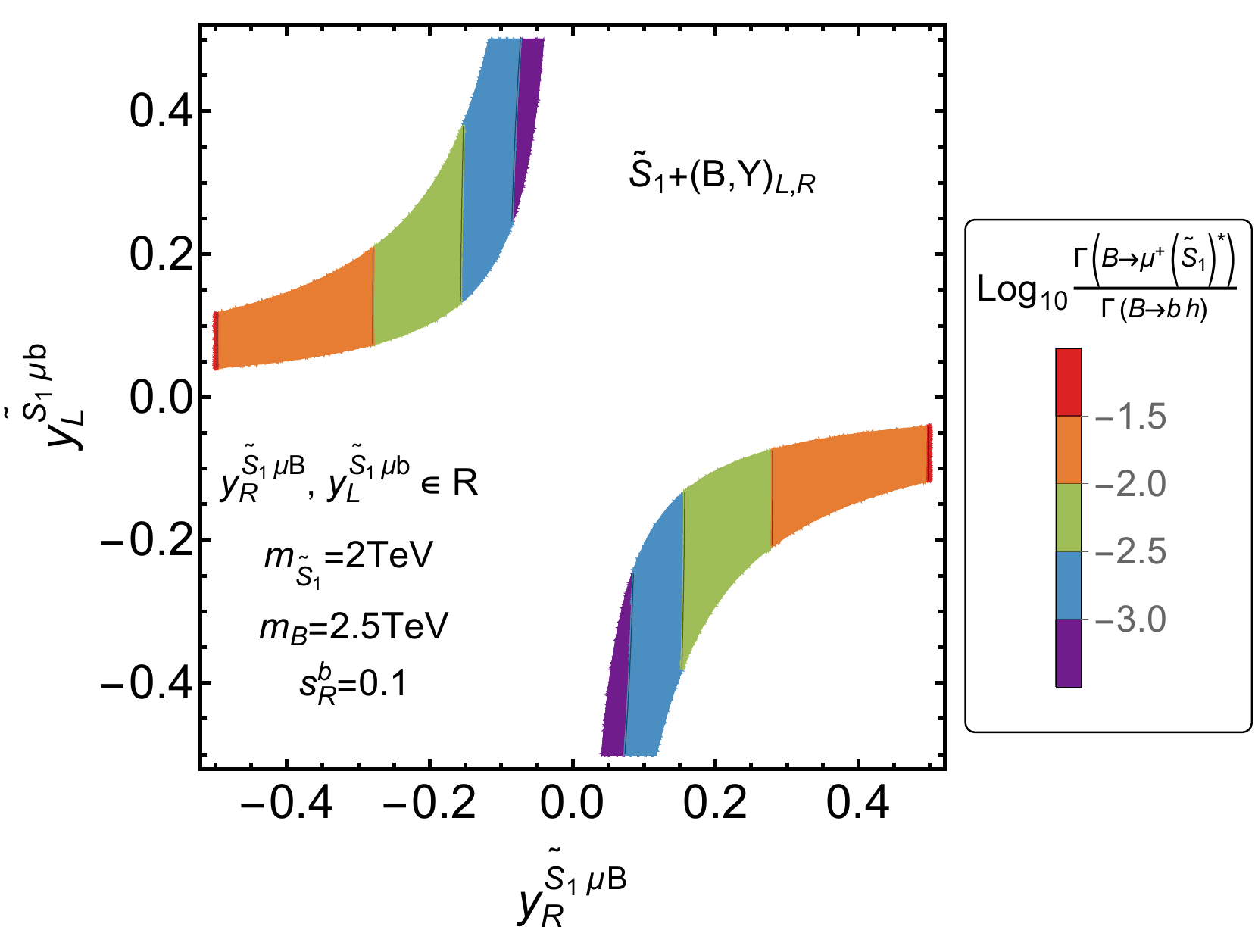}
\caption{The contour plots of $\log_{10}\frac{\Gamma(\widetilde{R}_2^{2/3}\rightarrow \mu^+B)}{\Gamma(\widetilde{R}_2^{2/3}\rightarrow \mu^+b)}$ (upper left), $\log_{10}\frac{\Gamma(\widetilde{S}_1\rightarrow \mu^+\bar{B})}{\Gamma(\widetilde{S}_1\rightarrow \mu^+\bar{b})}$ (lower left), $\log_{10}\frac{\Gamma(B\rightarrow\mu^-\widetilde{R}_2^{2/3})}{\Gamma(B\rightarrow bh)}$ (upper right), and $\log_{10}\frac{\Gamma(B\rightarrow\mu^+(\widetilde{S}_1)^{\ast})}{\Gamma(B\rightarrow bh)}$ (lower right), where the colored regions are allowed by the $(g-2)_{\mu}$ at $2\sigma$ CL. For the LQ decay, we choose $m_B=1.5\mr{TeV}$ and $m_{\mr{LQ}}=2\mr{TeV}$. For the VLQ decay, we choose $m_B=2.5\mr{TeV}$ and $m_{\mr{LQ}}=2\mr{TeV}$.}\label{fig:LQ+BY:decay}
\end{center}
\end{figure}

To estimate the effects of new decay channels, we will compare the ratios of new partial decay widths to the tradition ones. Because of gauge symmetry, the different partial decay widths can be correlated. Then, we choose the following four ratios:
\begin{align}
&\frac{\Gamma(\widetilde{R}_2^{2/3}\rightarrow \mu^+B)}{\Gamma(\widetilde{R}_2^{2/3}\rightarrow \mu^+b)}\sim\frac{|y_L^{\widetilde{R}_2\mu B}|^2}{|y_R^{\widetilde{R}_2\mu b}|^2},\quad\frac{\Gamma(\widetilde{S}_1\rightarrow \mu^+\bar{B})}{\Gamma(\widetilde{S}_1\rightarrow \mu^+\bar{b})}\sim\frac{|y_R^{\widetilde{S}_1\mu B}|^2}{|y_L^{\widetilde{S}_1\mu b}|^2},\nonumber\\
&\frac{\Gamma(B\rightarrow\mu^-\widetilde{R}_2^{2/3})}{\Gamma(B\rightarrow bh)}\sim\frac{v^2|y_L^{\widetilde{R}_2\mu B}|^2}{m_B^2(s_R^b)^2},\quad\frac{\Gamma(B\rightarrow\mu^+(\widetilde{S}_1)^{\ast})}{\Gamma(B\rightarrow bh)}\sim\frac{v^2|y_R^{\widetilde{S}_1\mu B}|^2}{m_B^2(s_R^b)^2}.
\end{align}
In Fig. \ref{fig:LQ+BY:decay}, we show the contour plots of above four ratios under the consideration of $(g-2)_{\mu}$ constraints. In these plots, we have included the full contributions.
We find that the new LQ decay channels can become important for larger $|y_L^{\widetilde{R}_2\mu B}|$ in the $\widetilde{R}_2+(B,Y)_{L,R}$ model and $|y_R^{\widetilde{S}_1\mu B}|$ in the $\widetilde{S}_1+(B,Y)_{L,R}$ model. As to the VLQ decay, the importance of new decay channels depends on the $s_R^b$ sensitively. For $s_R^b=0.1$ and $m_B=2.5\mr{TeV}$, the new VLQ decay channels are less significant. For smaller $s_R^b$, the new VLQ decay channels can play the important role.

For the LQ and VLQ production at hadron colliders, there are pair and single production channels, which are very sensitive to the LQ and VLQ masses. We can adopt the \textbf{FeynRules} \cite{Alloul:2013bka} to generate the model files and compute the cross sections with \textbf{MadGraph5${}_{-}$aMC@NLO} \cite{Alwall:2014hca}. For the 2 TeV scale LQ pair production \cite{Blumlein:1996qp, Diaz:2017lit, Dorsner:2018ynv}, the cross section can be $\sim0.01\mr{fb}$ at 13 TeV LHC. For the 1.5 TeV and 2.5 TeV scale VLQ pair production \cite{Aguilar-Saavedra:2009xmz, Matsedonskyi:2014mna, Fuks:2016ftf}, the cross section can be $\sim2\mr{fb}$ and $\sim0.01\mr{fb}$ at 13 TeV LHC. For the single LQ and VLQ production channels, they depend on the electroweak couplings \cite{Buonocore:2020erb, Buonocore:2022msy, Aguilar-Saavedra:2013qpa}. In the parameter space of large LQ Yukawa couplings, the single LQ production can be important, which may give some constraints at HL-LHC. To generate enough events, higher energy hadron colliders, for example 27 TeV and 100 TeV, can be necessary. Besides the collider direct search, there can be some indirect footprints, for example, $B$ physics related decay modes $\Upsilon\rightarrow\mu^+\mu^-,\nu\bar{\nu}\gamma$. If we consider more complex flavour structure (say, turn on the $\mr{LQ}\mu s$ interaction), it can also affect the $B\rightarrow K\mu^+\mu^-$ channel. Here, we will not study these detailed phenomenology.

\section{Summary and conclusions}

In this paper, we study the scalar LQ and VLQ extended models to explain the $(g-2)_{\mu}$ anomaly. Then, we find two new models $\widetilde{R}_2/\widetilde{S}_1+(B,Y)_{L,R}$, which can lead to the $B$ quark chiral enhancements because of the bottom and bottom partner mixing. During the numerical analysis, we consider two scenarios $m_{\mr{LQ}}>m_B$ and $m_{\mr{LQ}}<m_B$. After considering the experimental constraints, we choose relative light masses, which are adopted to be $(m_B,m_{\mr{LQ}})=(1.5\mr{TeV},2\mr{TeV})$ for the first scenario and $(m_B,m_{\mr{LQ}})=(2.5\mr{TeV},2\mr{TeV})$ for the second scenario. In the $\widetilde{R}_2+(B,Y)_{L,R}$ model, the $|\mr{Re}[y_L^{\widetilde{R}_2\mu B}(y_R^{\widetilde{R}_2\mu b})^\ast]|$ is bounded to be $\mc{O}(1)$, because $f_{LR}^{\widetilde{R}_2\mu b}(x)$ vanishes accidentally as $m_B= m_{\widetilde{R}_2}$. While, we can expect smaller $|\mr{Re}[y_L^{\widetilde{R}_2\mu B}(y_R^{\widetilde{R}_2\mu b})^\ast]|$ for largely splitted $m_{\widetilde{R}_2}$ and $m_B$.
In the $\widetilde{S}_1+(B,Y)_{L,R}$ model, the $\mr{Re}[y_R^{\widetilde{S}_1\mu B}(y_L^{\widetilde{S}_1\mu b})^\ast]$ is bounded to be the range $(-0.06, -0.02)$ at $2\sigma$ CL if $s_R^b=0.1$.
 
Under the constraints from $(g-2)_{\mu}$, we propose new LQ and VLQ search channels. In the scenario $m_{\mr{LQ}}>m_B$, there are new LQ decay channels: $\widetilde{R}_2^{2/3}\rightarrow \mu^+B$, $\widetilde{R}_2^{-1/3}\rightarrow \mu^+Y$, and $\widetilde{S}_1\rightarrow \mu^+\bar{B},\nu_L\bar{Y}$. For larger $y_L^{\widetilde{R}_2\mu B}$ and $y_R^{\widetilde{S}_1\mu B}$, it is important to take into account these decay channels. In the scenario $m_{\mr{LQ}}<m_B$, there are new VLQ decay channels: $B\rightarrow \mu^-\widetilde{R}_2^{2/3},\mu^+(\widetilde{S}_1)^{\ast}$ and $Y\rightarrow \mu^-\widetilde{R}_2^{-1/3},\nu_L(\widetilde{S}_1)^{\ast}$. For $s_R^b=0.1$, these channels are negligible compared to the traditional $B\rightarrow bZ,bh$ and $Y\rightarrow bW^-$ channels. For smaller $s_R^b$, these new VLQ decay channels can also become important.\\

\textit{Note added:} In paper \cite{Chowdhury:2022dps}, the authors study the model with $\widetilde{R}_2$, $S_3$, and $(B,Y)_{L,R}$. In their work, they do not consider the bottom and $B$ quark mixing, and the chiral enhancements are produced through the $\widetilde{R}_2$ and $S_3$ mixing. In paper \cite{Bigaran:2019bqv}, the authors explain the $(g-2)_{\mu}$ and $B$ physics anomalies in the $S_1+(B,Y)_{L,R}$ model.
\begin{acknowledgments}
This research was supported by an appointment to the Young Scientist Training Program at the APCTP through the Science and Technology Promotion Fund and Lottery Fund of the Korean Government. This was also supported by the Korean Local Governments-Gyeongsangbuk-do Province and Pohang City.
\end{acknowledgments}
\appendix

\section{LQ decay width formulae}\label{app:decay:LQ}
When the $\widetilde{R}_2$ masses are degenerate, there are no gauge boson decay channels such as $\widetilde{R}_2^{2/3}\rightarrow\widetilde{R}_2^{-1/3}W^+$. For the $\widetilde{R}_2^{2/3}$ to $\mu^+b$ and $\mu^+B$ decay channels, the widths are calculated as
\begin{align}
&\Gamma(\widetilde{R}_2^{2/3}\rightarrow\mu^+b)=\frac{m_{\widetilde{R}_2}}{16\pi}\sqrt{(1-\frac{m_{\mu}^2+m_b^2}{m_{\widetilde{R}_2}^2})^2-\frac{4m_{\mu}^2m_b^2}{m_{\widetilde{R}_2}^4}}~\cdot\nonumber\\
&\Big\{(1-\frac{m_{\mu}^2+m_b^2}{m_{\widetilde{R}_2}^2})[|y_L^{\widetilde{R}_2\mu B}|^2(s_L^b)^2+|y_R^{\widetilde{R}_2\mu b}|^2(c_R^b)^2]+\frac{4m_{\mu}m_b}{m_{\widetilde{R}_2}^2}s_L^bc_R^b\mr{Re}[y_L^{\widetilde{R}_2\mu B}(y_R^{\widetilde{R}_2\mu b})^\ast]\Big\},\nonumber\\
&\Gamma(\widetilde{R}_2^{2/3}\rightarrow\mu^+B)=\frac{m_{\widetilde{R}_2}}{16\pi}\sqrt{(1-\frac{m_{\mu}^2+m_B^2}{m_{\widetilde{R}_2}^2})^2-\frac{4m_{\mu}^2m_B^2}{m_{\widetilde{R}_2}^4}}~\cdot\nonumber\\
&\Big\{(1-\frac{m_{\mu}^2+m_B^2}{m_{\widetilde{R}_2}^2})[|y_L^{\widetilde{R}_2\mu B}|^2(c_L^b)^2+|y_R^{\widetilde{R}_2\mu b}|^2(s_R^b)^2]-\frac{4m_{\mu}m_B}{m_{\widetilde{R}_2}^2}c_L^bs_R^b\mr{Re}[y_L^{\widetilde{R}_2\mu B}(y_R^{\widetilde{R}_2\mu b})^\ast]\Big\}.
\end{align}
For the $\widetilde{R}_2^{-1/3}$ to $\mu^+Y,\nu_Lb,\nu_LB$ decay channels, the widths are calculated as
\begin{align}
&\Gamma(\widetilde{R}_2^{-1/3}\rightarrow\mu^+Y)=\frac{m_{\widetilde{R}_2}}{16\pi}\sqrt{(1-\frac{m_{\mu}^2+m_Y^2}{m_{\widetilde{R}_2}^2})^2-\frac{4m_{\mu}^2m_Y^2}{m_{\widetilde{R}_2}^4}}(1-\frac{m_{\mu}^2+m_Y^2}{m_{\widetilde{R}_2}^2})|y_L^{\widetilde{R}_2\mu B}|^2,\nonumber\\
&\Gamma(\widetilde{R}_2^{-1/3}\rightarrow\nu_Lb)=\frac{m_{\widetilde{R}_2}}{16\pi}(1-\frac{m_b^2}{m_{\widetilde{R}_2}^2})^2|y_R^{\widetilde{R}_2\mu b}|^2(c_R^b)^2,\nonumber\\
&\Gamma(\widetilde{R}_2^{-1/3}\rightarrow\nu_LB)=\frac{m_{\widetilde{R}_2}}{16\pi}(1-\frac{m_B^2}{m_{\widetilde{R}_2}^2})^2|y_R^{\widetilde{R}_2\mu b}|^2(s_R^b)^2.
\end{align}
Considering $m_{\mu},m_b\ll m_B$ and $\theta_{L,R}^b\ll1$, we have the following approximations:
\begin{align}
&\Gamma(\widetilde{R}_2^{2/3}\rightarrow\mu^+B)\approx\Gamma(\widetilde{R}_2^{-1/3}\rightarrow\mu^+Y)\approx\frac{m_{\widetilde{R}_2}}{16\pi}(1-\frac{m_B^2}{m_{\widetilde{R}_2}^2})^2|y_L^{\widetilde{R}_2\mu B}|^2,\nonumber\\
&\Gamma(\widetilde{R}_2^{2/3}\rightarrow\mu^+b)\approx\Gamma(\widetilde{R}_2^{-1/3}\rightarrow\nu_Lb)\approx\frac{m_{\widetilde{R}_2}}{16\pi}|y_R^{\widetilde{R}_2\mu b}|^2.
\end{align}

For the $\widetilde{S}_1$ to $\mu^+\bar{b},\mu^+\bar{B},\nu_L\bar{Y}$ decay channels, the widths are calculated as
\begin{align}
&\Gamma(\widetilde{S}_1\rightarrow\mu^+\bar{b})=\frac{m_{\widetilde{S}_1}}{16\pi}\sqrt{(1-\frac{m_{\mu}^2+m_b^2}{m_{\widetilde{S}_1}^2})^2-\frac{4m_{\mu}^2m_b^2}{m_{\widetilde{S}_1}^4}}~\cdot\nonumber\\
&\Big\{(1-\frac{m_{\mu}^2+m_b^2}{m_{\widetilde{S}_1}^2})[|y_R^{\widetilde{S}_1\mu B}|^2(s_L^b)^2+|y_L^{\widetilde{S}_1\mu b}|^2(c_R^b)^2]+\frac{4m_{\mu}m_b}{m_{\widetilde{S}_1}^2}s_L^bc_R^b\mr{Re}[y_R^{\widetilde{S}_1\mu B}(y_L^{\widetilde{S}_1\mu b})^\ast]\Big\},\nonumber\\
&\Gamma(\widetilde{S}_1\rightarrow\mu^+\bar{B})=\frac{m_{\widetilde{S}_1}}{16\pi}\sqrt{(1-\frac{m_{\mu}^2+m_B^2}{m_{\widetilde{S}_1}^2})^2-\frac{4m_{\mu}^2m_B^2}{m_{\widetilde{S}_1}^4}}~\cdot\nonumber\\
&\Big\{(1-\frac{m_{\mu}^2+m_B^2}{m_{\widetilde{S}_1}^2})[|y_R^{\widetilde{S}_1\mu B}|^2(c_L^b)^2+|y_L^{\widetilde{S}_1\mu b}|^2(s_R^b)^2]-\frac{4m_{\mu}m_B}{m_{\widetilde{S}_1}^2}c_L^bs_R^b\mr{Re}[y_R^{\widetilde{S}_1\mu B}(y_L^{\widetilde{S}_1\mu b})^\ast]\Big\},\nonumber\\
&\Gamma(\widetilde{S}_1\rightarrow\nu_L\bar{Y})=\frac{m_{\widetilde{S}_1}}{16\pi}(1-\frac{m_Y^2}{m_{\widetilde{S}_1}^2})^2|y_R^{\widetilde{S}_1\mu B}|^2.
\end{align}
Considering $m_{\mu},m_b\ll m_B$ and $\theta_{L,R}^b\ll1$, we have the following approximations:
\begin{align}
&\Gamma(\widetilde{S}_1\rightarrow\mu^+\bar{B})\approx\Gamma(\widetilde{S}_1\rightarrow\nu_L\bar{Y})\approx\frac{m_{\widetilde{S}_1}}{16\pi}(1-\frac{m_B^2}{m_{\widetilde{S}_1}^2})^2|y_R^{\widetilde{S}_1\mu B}|^2,\nonumber\\
&\Gamma(\widetilde{S}_1\rightarrow\mu^+\bar{b})\approx\frac{m_{\widetilde{S}_1}}{16\pi}|y_L^{\widetilde{S}_1\mu b}|^2.
\end{align}
\section{VLQ decay width formulae}\label{app:decay:VLQ}
If $m_B\sim\mr{TeV}$ and $\theta_R^b\ll0.1$, we have $m_B-m_Y\approx m_B(s_R^b)^2/2\lesssim5\mr{GeV}$, which leads to the kinematic prohibition of some decay channels. For the $Y\rightarrow bW^-$ decay channel, the width is calculated as
\begin{align}
&\Gamma(Y\rightarrow bW^-)=\frac{g^2}{64\pi m_Y}\sqrt{(1-\frac{m_b^2+m_W^2}{m_Y^2})^2-\frac{4m_b^2m_W^2}{m_Y^4}}\cdot\nonumber\\
&\Big\{[(s_L^b)^2+(s_R^b)^2]\frac{(m_Y^2-m_b^2)^2+m_W^2(m_Y^2+m_b^2)-2m_W^4}{m_W^2}-12m_Ym_bs_L^bs_R^b\Big\}.
\end{align}
For the $B\rightarrow bZ,bh,tW^-$ decay channels, the widths are calculated as
\begin{align}
&\Gamma(B\rightarrow bh)=\frac{m_B}{32\pi}\sqrt{(1-\frac{m_b^2+m_h^2}{m_B^2})^2-\frac{4m_b^2m_h^2}{m_B^4}}\Big[(1+\frac{m_b^2-m_h^2}{m_B^2})\frac{m_b^2+m_B^2}{v^2}+4\frac{m_b^2}{v^2}\big)\Big](s_R^b)^2(c_R^b)^2,\nonumber\\
&\Gamma(B\rightarrow bZ)=\frac{g^2}{32\pi c_W^2m_B}\sqrt{(1-\frac{m_b^2+m_Z^2}{m_B^2})^2-\frac{4m_b^2m_Z^2}{m_B^4}}\cdot\nonumber\\
&\Big\{[(s_L^bc_L^b)^2+\frac{(s_R^bc_R^b)^2}{4}]\frac{(m_B^2-m_b^2)^2+m_Z^2(m_B^2+m_b^2)-2m_Z^4}{m_Z^2}-6m_Bm_bs_L^bs_R^bc_L^bc_R^b\Big\},\nonumber\\
&\Gamma(B\rightarrow tW^-)=\frac{g^2(s_L^b)^2}{64\pi}\sqrt{(1-\frac{m_t^2+m_W^2}{m_B^2})^2-\frac{4m_t^2m_W^2}{m_B^4}}\frac{(m_B^2-m_t^2)^2+m_W^2(m_B^2+m_t^2)-2m_W^4}{m_W^2m_B}.
\end{align}
Considering $m_b,m_t,m_Z,m_W\ll m_B$ and $\theta_{L,R}^b\ll1$, we have the following approximations:
\begin{align}
&\Gamma(B\rightarrow bZ)\approx\Gamma(B\rightarrow bh)\approx\frac{1}{2}\Gamma(Y\rightarrow bW^-)\approx\frac{m_B^3}{32\pi v^2}(s_R^b)^2,\nonumber\\
&\Gamma(B\rightarrow tW^-)\approx\frac{m_b^2m_B}{16\pi v^2}(s_R^b)^2.
\end{align}

In the $\widetilde{R}_2+(B,Y)_{L,R}$ model, the VLQ can also decay into $\widetilde{R}_2$ final state. For the $Y\rightarrow \mu^-\widetilde{R}_2^{-1/3}$ decay channel, the width is calculated as
\begin{align}
&\Gamma(Y\rightarrow\mu^-\widetilde{R}_2^{-1/3})=\frac{m_Y}{32\pi}\sqrt{(1-\frac{m_{\mu}^2+m_{\widetilde{R}_2}^2}{m_Y^2})^2-\frac{4m_{\mu}^2m_{\widetilde{R}_2}^2}{m_Y^4}}(1+\frac{m_{\mu}^2-m_{\widetilde{R}_2}^2}{m_Y^2})|y_L^{\widetilde{R}_2\mu B}|^2.
\end{align}
For the $B\rightarrow \mu^-\widetilde{R}_2^{2/3},\nu_L\widetilde{R}_2^{-1/3}$ decay channels, the widths are calculated as
\begin{align}
&\Gamma(B\rightarrow\mu^-\widetilde{R}_2^{2/3})=\frac{m_B}{32\pi}\sqrt{(1-\frac{m_{\mu}^2+m_{\widetilde{R}_2}^2}{m_B^2})^2-\frac{4m_{\mu}^2m_{\widetilde{R}_2}^2}{m_B^4}}~\cdot\nonumber\\
&\Big\{(1+\frac{m_{\mu}^2-m_{\widetilde{R}_2}^2}{m_B^2})[|y_L^{\widetilde{R}_2\mu B}|^2(c_L^b)^2+|y_R^{\widetilde{R}_2\mu b}|^2(s_R^b)^2]+\frac{4m_{\mu}m_{\widetilde{R}_2}}{m_B^2}c_L^bs_R^b\mr{Re}[y_L^{\widetilde{R}_2\mu B}(y_R^{\widetilde{R}_2\mu b})^\ast]\Big\},\nonumber\\
&\Gamma(B\rightarrow\nu_L\widetilde{R}_2^{-1/3})=\frac{m_B}{32\pi}(1-\frac{m_{\widetilde{R}_2}^2}{m_B^2})^2|y_R^{\widetilde{R}_2\mu b}|^2(s_R^b)^2.
\end{align}
Considering $m_{\mu}\ll m_B$ and $\theta_{L,R}^b\ll1$, we have the following approximations:
\begin{align}
&\Gamma(B\rightarrow\mu^-\widetilde{R}_2^{2/3})\approx\Gamma(Y\rightarrow\mu^-\widetilde{R}_2^{-1/3})\approx\frac{m_B}{32\pi}(1-\frac{m_{\widetilde{R}_2}^2}{m_B^2})^2|y_L^{\widetilde{R}_2\mu B}|^2.
\end{align}

In the $\widetilde{S}_1+(B,Y)_{L,R}$ model, the VLQ can also decay into $\widetilde{S}_1$ final state. For the $Y\rightarrow \nu_L(\widetilde{S}_1)^{\ast}$ decay channel, the width is calculated as
\begin{align}
&\Gamma(Y\rightarrow\nu_L(\widetilde{S}_1)^{\ast})=\frac{m_Y}{32\pi}(1-\frac{m_{\widetilde{S}_1}^2}{m_Y^2})^2|y_R^{\widetilde{S}_1\mu B}|^2.
\end{align}
For the $B\rightarrow \mu^+(\widetilde{S}_1)^{\ast}$ decay channel, the width is calculated as
\begin{align}
&\Gamma(B\rightarrow\mu^+(\widetilde{S}_1)^{\ast})=\frac{m_B}{32\pi}\sqrt{(1-\frac{m_{\mu}^2+m_{\widetilde{S}_1}^2}{m_B^2})^2-\frac{4m_{\mu}^2m_{\widetilde{S}_1}^2}{m_B^4}}~\cdot\nonumber\\
&\Big\{(1+\frac{m_{\mu}^2-m_{\widetilde{S}_1}^2}{m_B^2})[|y_R^{\widetilde{S}_1\mu B}|^2(c_L^b)^2+|y_L^{\widetilde{S}_1\mu b}|^2(s_R^b)^2]+\frac{4m_{\mu}}{m_B}c_L^bs_R^b\mr{Re}[y_R^{\widetilde{S}_1\mu B}(y_L^{\widetilde{S}_1\mu b})^\ast]\Big\}.
\end{align}
Considering $m_{\mu}\ll m_B$ and $\theta_{L,R}^b\ll1$, we have the following approximations:
\begin{align}
\Gamma(B\rightarrow\mu^+(\widetilde{S}_1)^{\ast})\approx\Gamma(Y\rightarrow\nu_L(\widetilde{S}_1)^{\ast})\approx\frac{m_B}{32\pi}(1-\frac{m_{\widetilde{S}_1}^2}{m_B^2})^2|y_R^{\widetilde{S}_1\mu B}|^2.
\end{align}
\end{sloppypar}
\bibliography{muongm2-LQ-VLQB}
\end{document}